# High efficiency second harmonic generation in transverse orientation patterned gallium phosphide waveguides

ANTOINE LEMOINE[1], BRIEG LE CORRE[1,2], LISE MORICE[1], ABDELMOUNAÏM HAROURI[2], LUC LE GRATIET[2], GREGOIRE BEAUDOIN[2], JULIE LE POULIQUEN[1], KARINE TAVERNIER[1], ARNAUD GRISARD[3], SYLVAIN COMBRIE[3], BRUNO GERARD[4], CHARLES CORNET[1], CHRISTOPHE LEVALLOIS[1], YANNICK DUMEIGE[1], KONSTANTINOS PANTZAS[2], ISABELLE SAGNES[2], YOAN LEGER[1,*]

[1]*Univ Rennes, INSA Rennes, CNRS, Institut FOTON - UMR 6082, F-35000 Rennes, France*
[2]*Centre de Nanosciences et de Nanotechnologie, CNRS, 91120 Palaiseau, France*
[3]*Thales Research and Technology, 91767 Palaiseau, France*
[4]*III-V Lab, 91767 Palaiseau, France*
**yoan.leger@insa-rennes.fr*

**Abstract:** Achieving high conversion efficiencies in second-order nonlinear optical processes is a key challenge in integrated photonics for both classical and quantum applications. This paper presents the first demonstration of Transverse Orientation-Patterned gallium phosphide (TOP-GaP) waveguides showing high-efficiency second harmonic generation. In such devices, first order modal phase matching is unlocked and optimized through the inversion of the nonlinear susceptibility along the vertical direction. We discuss here the theory behind modal phase matching in TOP structures, describe the fabrication process, and present linear and nonlinear characterizations of the TOP-GaP waveguides.

## 1. Introduction

Reaching high efficiencies in second order nonlinear photonic integrated devices is the key to unlock numerous applications in both classical [1],[2] and quantum light sources [3], [4] on chip, through processes such as second-harmonic generation (SHG), sum-frequency generation, and spontaneous parametric down-conversion. High efficiencies frequency conversion can be achieved through two main approaches: reducing optical losses, which depends on the technological maturity of the photonic platform, and optimizing compensation for the significant phase mismatch often associated with second-order nonlinear processes; however, the ideal strategy combines both methods. For the latter, the most effective approach so far to overcome this limitation has been the use of quasi-phase matching (QPM) condition. The first demonstrations were made in bulk materials such as lithium niobate (LN) [5] and on various III-V photonic platforms with epitaxial layers, including gallium arsenide (GaAs) [6], [7], gallium nitride (GaN) [8], [9] and gallium phosphide (GaP). While periodic poling has been successfully scaled down to photonic circuits for the LN platform, with record conversion efficiencies of 4600% /W/cm² [10], chip integration has been more challenging with III-Vs, where the first QPM waveguides were demonstrated with AlGaAs [11]. More recently, state-of-the-art conversion efficiency has been achieved with QPM in GaP [12]. Another approach to phase matching arose directly from photonic integration. Benefiting of the modal dispersion of multimode waveguides, the material dispersion can be compensated through the use of different transverse modes for input and output fields. This modal phase matching was successfully demonstrated in several platforms [13], [14], [15], [16] but to the detriment of the conversion efficiency, due to the poor nonlinear overlap between the transverse modes involved in the nonlinear process and the nonlinear susceptibility. In addition, modal phase matching introduces symmetry restrictions on the parity of the modes, limiting even more the potential of this approach. To tackle this issue, the engineering of the nonlinear susceptibility of a waveguide in the transverse plane has been proposed on several photonic platforms, revisiting the original proposal of Khurgin [17]. Modal symmetry modification using this pioneering approach is achieved by inhibiting the formation of second-harmonic mode lobes through localized cancellation of nonlinear susceptibility, thereby enforcing the correct symmetry in the SH mode. This has been demonstrated in material stacks such as LN/$TiO_2$ [18], as well as in materials with crystallographic defects, like GaP on silicon, having antiphase domains in a limited thickness within the sample [19]. Another similar strategy to unlock parity-forbidden processes involves the local reversing of the sign of the nonlinear coefficient, which has been successfully applied in epitaxial bilayers like GaN/AlN [20]. Recently, this method was proposed for LN [21], leading to the realization of resonant devices with high conversion efficiencies [22], [23]. The transfer of such an approach to the GaP photonic platform could lead to even larger conversion efficiencies at technological maturity, extend the spectral range of second order nonlinear processes to the mid-infrared and represent an alternative to bypass the technological challenge of orientation-patterned III-Vs waveguide fabrication.

In this work, we report on the fabrication of transverse orientation-patterned (TOP) GaP nanowaveguides and demonstrate high efficiency second harmonic generation from 1310nm to 655nm. We first detail the theory behind modal phase matching in TOP-structures and discuss its relevance compared to perfect phase matching. We then describe the fabrication process of TOP-GaP nanowaveguides, based on membrane thermo-compression bonding. Finally, linear and nonlinear optical characterizations are presented and the measured conversion efficiency, that benefits of both strong light confinement and large material nonlinear susceptibility, is compared to theoretical predictions and state-of-the art demonstrations in other platforms.

## 2. Impact of mode overlap on conversion efficiency in TOP-GaP

In a GaP waveguide, the theoretical conversion efficiency for SHG, accounting for propagation losses at the fundamental and SH frequencies reads as [12]:

$$\eta = \frac{2\omega^2}{\epsilon_0 c^3} \frac{L^2}{n_\omega^2 n_{2\omega}} \mathcal{H}\Gamma \tag{1}$$

with $\omega$ and $2\omega$ the angular frequencies of the fundamental and the SH modes, $n_\omega$ and $n_{2\omega}$ their respective effective refractive indices, $\epsilon_0$ the vacuum permittivity, $c$ the speed of light. The mismatch function $\mathcal{H}$ is:

$$\mathcal{H} = \frac{\left(\sinh\left(\frac{\Delta\alpha L}{2}\right)\cos\left(\frac{\Delta\beta L}{2}\right)\right)^2 + \left(\cosh\left(\frac{\Delta\alpha L}{2}\right)\sin\left(\frac{\Delta\beta L}{2}\right)\right)^2}{(\Delta\alpha^2 + \Delta\beta^2)L^2/4} \times e^{-\left(\frac{\alpha_{2\omega}}{2}+\alpha_\omega\right)L} \tag{2}$$

With $\Delta\alpha$ and $\Delta\beta$ the loss and phase mismatches. The loss mismatch is defined as $\Delta\alpha = \frac{\alpha_{2\omega}}{2} - \alpha_\omega$ with $\alpha_\omega$ and $\alpha_{2\omega}$ respectively the fundamental and SH propagation losses. The phase mismatch is defined as $\Delta\beta = \beta_{2\omega} - 2\beta_\omega$ with $\beta_\omega = \omega n_\omega/c$ and $\beta_{2\omega} = 2\omega n_{2\omega}/c$. Finally, the nonlinear overlap integral $\Gamma$ is:

$$\Gamma = \frac{\left(\iint d_{eff}\, E_\omega^2(x,y) E_{2\omega}^*(x,y)\mathrm{d}x\mathrm{d}y\right)^2}{\left(\iint |E_\omega\,(x,y)|^2\,\mathrm{d}x\mathrm{d}y\right)^2 \ \iint |E_{2\omega}(x,y)|^2 \mathrm{d}x\mathrm{d}y} \tag{3}$$

$d_{eff} = d_{14}(x,y)$ is the spatial distribution of the nonlinear coefficient in the waveguide whose maximal value is equal to $\pm$ 36.8 pm/V at 1310 nm [24]. Here $E_\omega$ represents the fundamental electric field, and $E_{2\omega}$ represents the SH electric field. In GaP, the zinc blende crystal symmetry imposes that a waveguide oriented along the [110] direction enables the interaction of two photons in TE polarization and one in TM polarization. This work focuses on the most common type I configuration for SHG, which involves two TE-polarized pump photons and one TM-polarized photon. Equation (3) shows that the nonlinear overlap vanishes for antisymmetric SH modes, imposing a selection rule on even SH modes for the nonlinear process. Engineering the spatial distribution of the nonlinear coefficient can lift this selection rule and optimize $\Gamma$ by providing $d_{14}$ with the same symmetry as $E_{2\omega}$. In contrast to LN where ferroelectric poling can apply, the reversal of $d_{14}$ requires a physical $\pi$ rotation of the lattice around a [110] axis. The waveguide architecture proposed here is therefore composed of two layers of GaP with opposite crystal orientation, enabling a change of the sign of the $d_{14}$ coefficient between the upper and lower halves of the waveguide as shown schematically in Fig. 1(a). This $d_{14}$ distribution unlocks SHG towards any vertical antisymmetric SH mode and in particular the first one, $TM_{10}$, for which the nonlinear overlap is the greatest. Taking advantage of the large refractive index of GaP, $\Gamma$ can also be optimized by decreasing the dimensions of the waveguide. To phase-match between a $TE_{00}$ pump at 1310 nm and a $TM_{10}$ SH signal at 655nm, waveguides with a total thickness of 325 nm and a width of 820 nm can be used as presented in Fig. 1(b). In this configuration, $\Gamma$ can be calculated to reach 80% of the nonlinear overlap that a perfect phase matching could reach for these dimensions, that is to say between a $TE_{00}$ pump and a $TM_{00}$ SH mode. This value will later be taken into account when calculating the theoretical conversion efficiency based on propagation losses measured in the fabricated TOP-GaP waveguide.

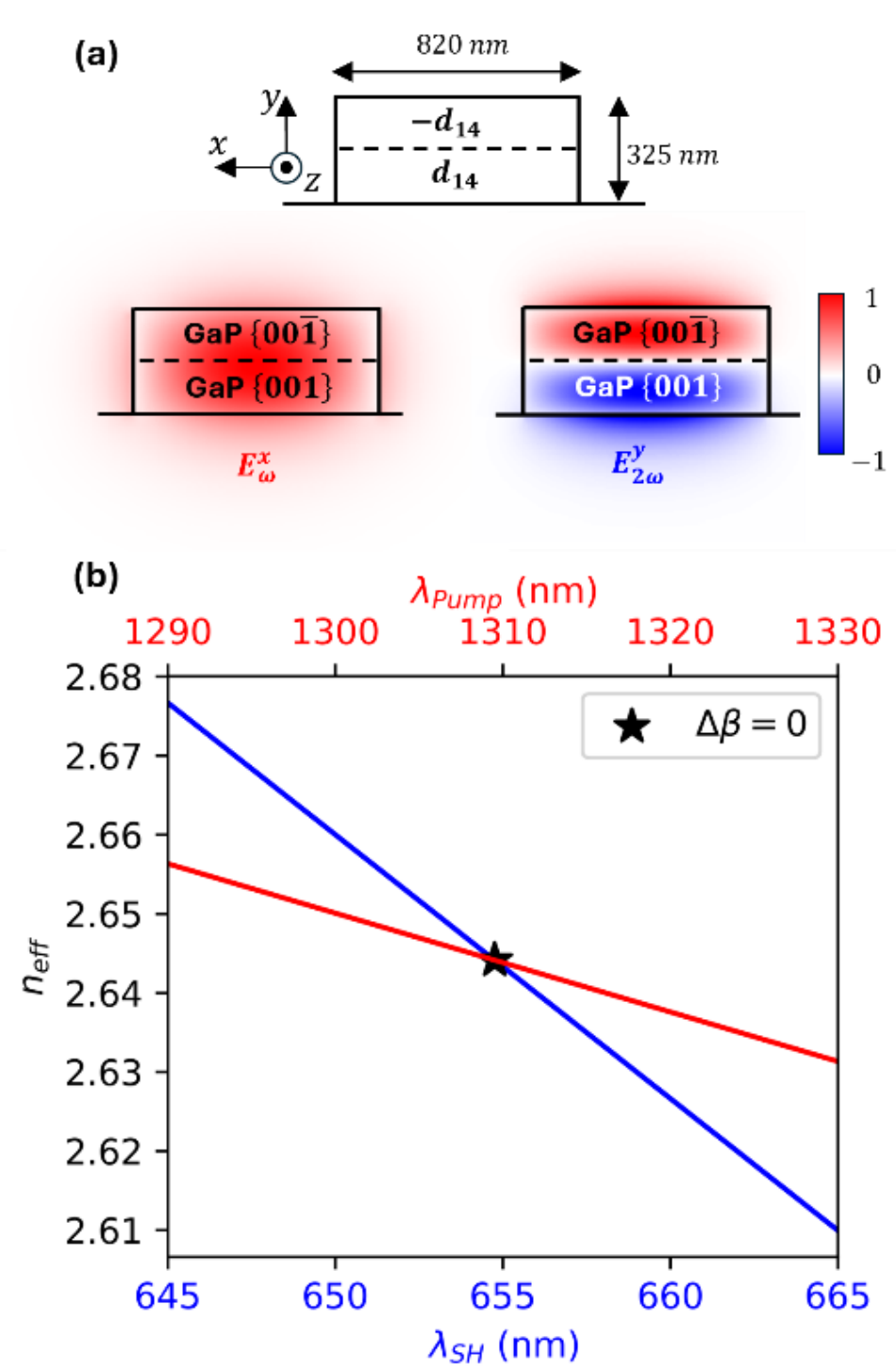

Figure 1 a) Schematic view of a TOP-GaP waveguide cross section with nonlinear coefficient spatial distribution and the electric-field components of interest. b) Simulated mode dispersion corresponding to $TE_{00}$ (pump) and $TM_{10}$ (SH). Phase matching occurs close to 1310 nm.

## 3. Fabrication process

The fabrication workflow of TOP-GaP nanowaveguides is presented in Fig. 2(a). The first steps of the TOP- GaP waveguide fabrication process are borrowed from the technology of OP-GaP seeds described in [12]. Thin GaP layers are first grown at the same thickness on an {001}-orientated GaAs substrate by metal-organic vapor-phase epitaxy (MOVPE). Then the two GaP layers are brought together and bonded head to tail along their respective (001) crystalline plane using thermal-compression bonding. The top GaP layer, which has been flipped by 180° around [110] then presents a reversed crystalline orientation ($\{00\bar{1}\}$-oriented) with respect to the one of the bottom GaP layer ({001}-oriented). The top substrate is then etched by using a $H_2SO_4$:$H_2O_2$:$H_2O$ solution which selectively etch GaAs on GaP. Contrary to Ref. [10], vertical optical confinement is obtained by transferring the TOP GaP bilayer membrane onto insulator. $SiO_2$ is first deposited on both the GaP stack and the Si substrate by PECVD. Then benzocyclobutene (BCB) is spined onto the host Si substrate and the two parts are bonded together by polymer bonding [25]. As before, the remaining GaAs substrate is removed using a $H_2SO_4$:$H_2O_2$:$H_2O$ solution, forming a GaP membrane onto insulator. Finally, the TOP-GaP waveguides are patterned by electronic lithography and induced coupled plasma (ICP) etching and cleaved in 1.3 mm waveguides. As shown in Fig. 2(b), the resist, with refractive index close to silica, is kept on top of the waveguides to ensure a good vertical symmetry of the mode profiles and limits the impact of roughness originating from the GaP/GaAs interface. On the SEM image, a slight contrast on the side of the waveguide can be observed between the {001} and $\{00\bar{1}\}$-oriented layers. The sidewall angle measured in Fig. 2(b) with tilt and rotation correction, is 4° from the growth direction, indicating that TE/TM coupling can be neglected in these waveguides. The sidewall roughness is estimated to 1 nm RMS by SEM measurements.

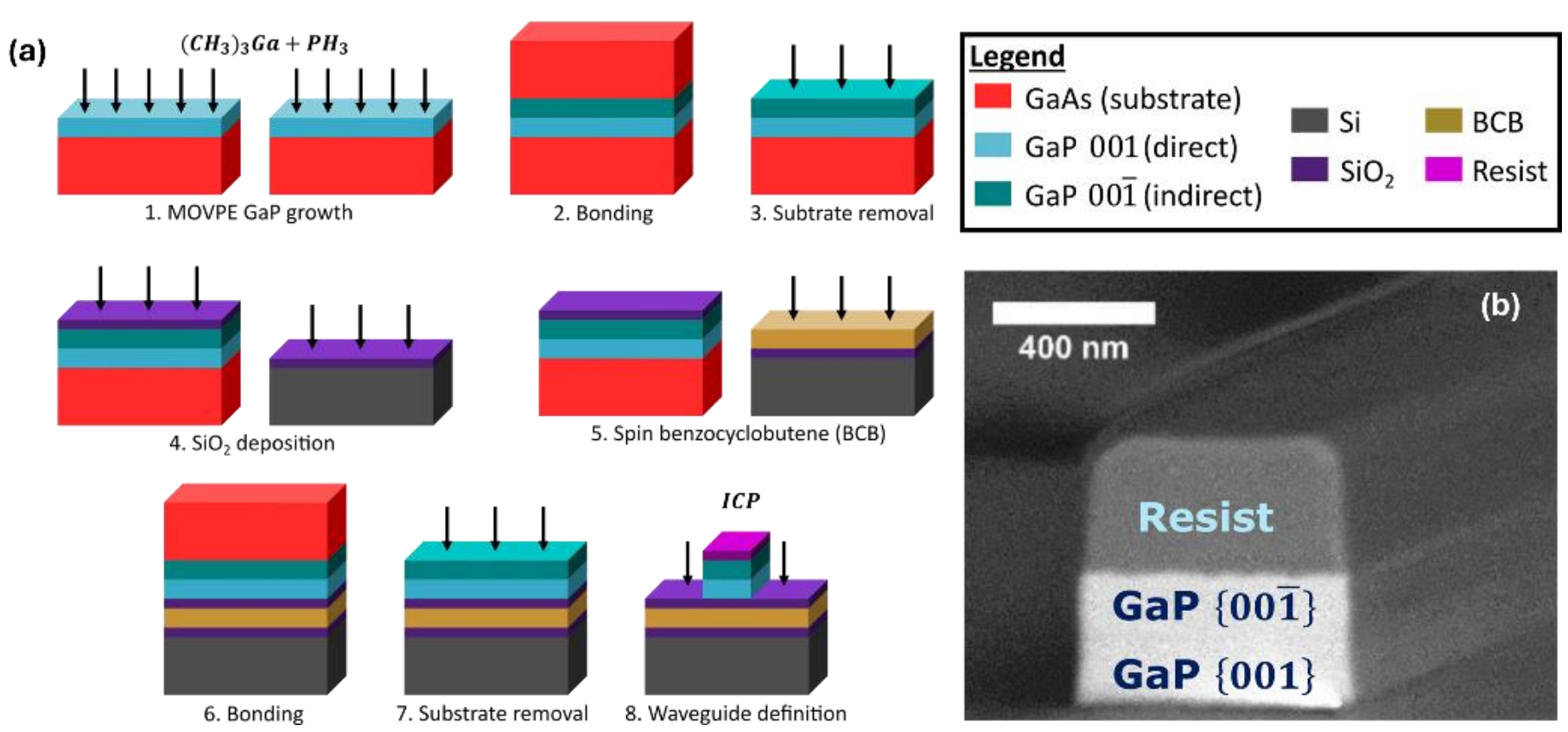

Figure 2 a) Fabrication steps of TOP-GaP waveguides. b) Scanning electron microscope (SEM) image of TOP-GaP cleaved waveguide.

## 4. Linear and nonlinear characterization

### *4.1 Propagation loss measurements*

To quantify the propagation losses at both the pump and SH wavelengths, two complementary methods were employed. First, Fabry-Perot (FP) fringe contrast measurements were used to evaluate the propagation losses in TOP-GaP waveguides at the pump wavelength [26]. These measurements were conducted with a tunable Littman-Metcalf laser operating in the O-band (Yenista T100S-HP), with polarization control to specifically address TE modes. Free-space butt-coupling was achieved using a x20 magnification microscope objective optimized for IR with a 0.4 numerical aperture, and data acquisition was carried out using germanium photodiodes. The contrast in the transmission spectra, presented in Fig. 3(a) is directly related to cavity losses, allowing us to estimate average propagation losses of approximately 40 dB/cm at the pump wavelength (see Fig. 3(b)). The modulation observed on the transmission spectrum is inferred to localized defects such as resist peeling or defects at the bonding interfaces along the waveguide. At the second harmonic wavelength, propagation losses were measured by analyzing surface light scattering of the propagating mode injected using a 635 nm laser diode with a microscope setup equipped with a scientific camera (Fig. 3(c)). By tracking the signal's evolution as a function of propagation length in the images, we estimated the SH propagation losses to be around 33 dB/cm at the SH wavelength (Fig. 3(d)). Note that this value applies mostly to the $TM_{00}$ mode addressed with our injection setup. These loss values are consistent with those reported in the literature for waveguides based on GaP grown on GaAs [13], [27] and are mainly attributed to surface roughness on the upper and lower interfaces of the guides, corresponding to the GaP/GaAs interface with RMS roughness of 2 nm. This roughness component being larger than the sidewall one, TE modes are expected to suffer larger losses than TM ones,[28] as experimentally observed. On the light scattering image, point like scattering coming from air gaps at the bonding interface can still clearly be observed in most waveguides. Thus, one may assume smaller losses for the $TM_{10}$ mode presenting a field node at the bonding interface plane.

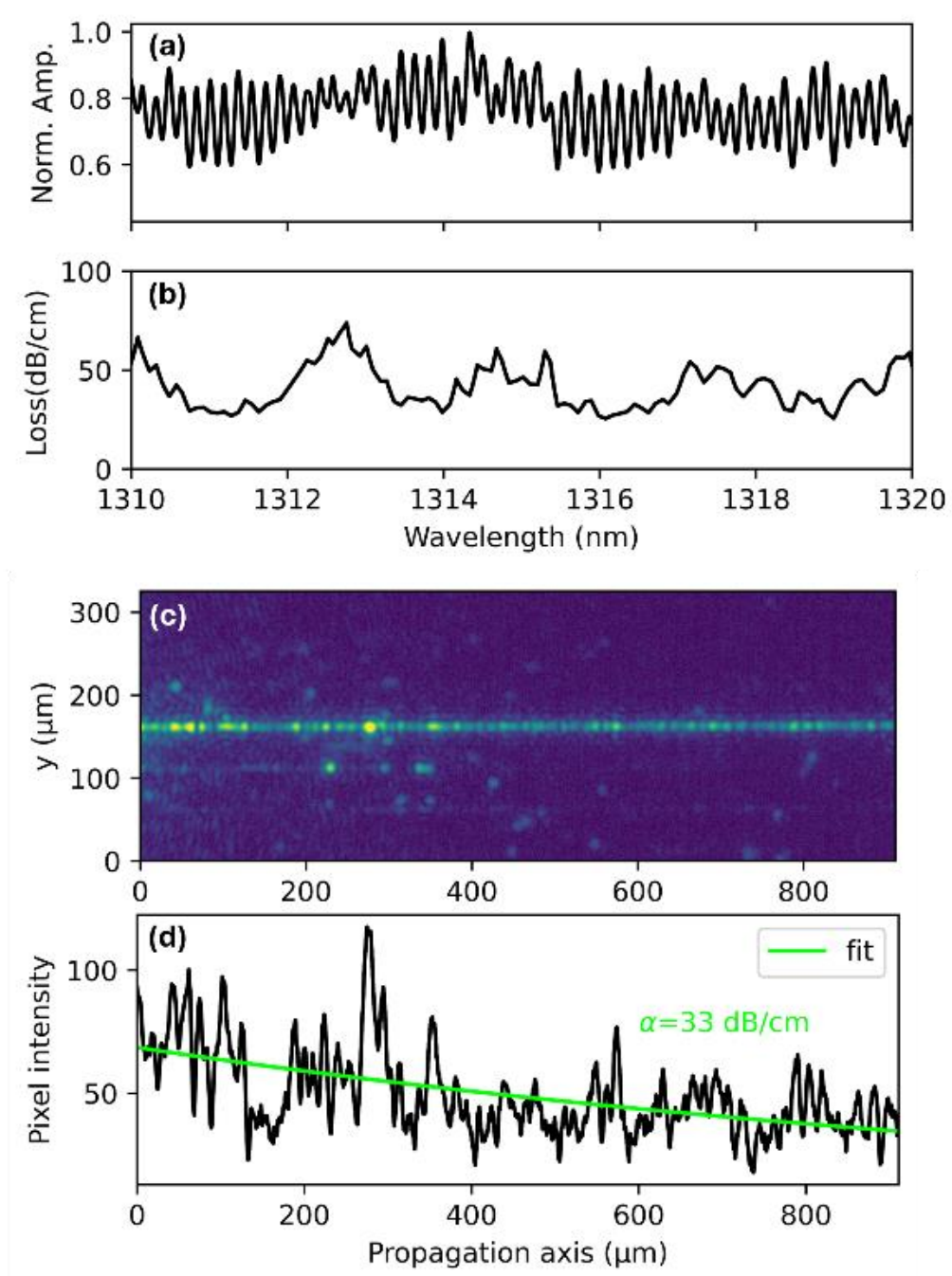

Figure 3 (a) Fabry-Perot fringes of the cavity formed by the TOP-GaP waveguides with cleaved facets in the O-band. (b) Optical propagation losses obtained from contrast analysis of the transmission spectrum shown in (a). (c) Top view image of the light scattered from the propagating mode in a TOP-GaP waveguide at 635 nm. (d) Line profile of (c) for loss evaluation.

### *4.2 Nonlinear characterization*

Nonlinear measurements were conducted using a modified version of the original experimental setup. To optimize light collection from the highly diverging $TM_{10}$ mode, the detection objective was replaced with a microscope objective optimized for the visible, with higher numerical aperture of 0.8. Furthermore, a beam splitter was added before the injection objective, enabling continuous monitoring of pump power during wavelength scans around the phase-matching condition. A shortpass filter at 700 nm was also added in free space after the collection objective in order to filter out the residual pump. SH light power was finally measured with a silicon photodiode connected to a power meter. Conversion efficiency is then calculated as $\eta = P_{2\omega}(L)/P_{\omega}^2(0)$. $P_{\omega}(0)$ is the amount of power injected into the waveguide, calculated from the power measured in front of the injection microscope objective and accounting for the transmission of the microscope objective, the butt coupling coefficient and the Fresnel reflection coefficient for the $TE_{00}$ mode at the entrance facet between air and GaP. $P_{2\omega}(L)$ is the second harmonic power at the end of the waveguide, deduced from the transmission of the optics between the detector and the ending facet. Finally, only 6% of the SH signal is collected due to the far-field distribution of the SH mode. Detail of calculation of $P_{\omega}^2(0)$ and $P_{2\omega}(L)$, derived from the measurement of the collected signal and laser power, are provided in the Supplementary Information.

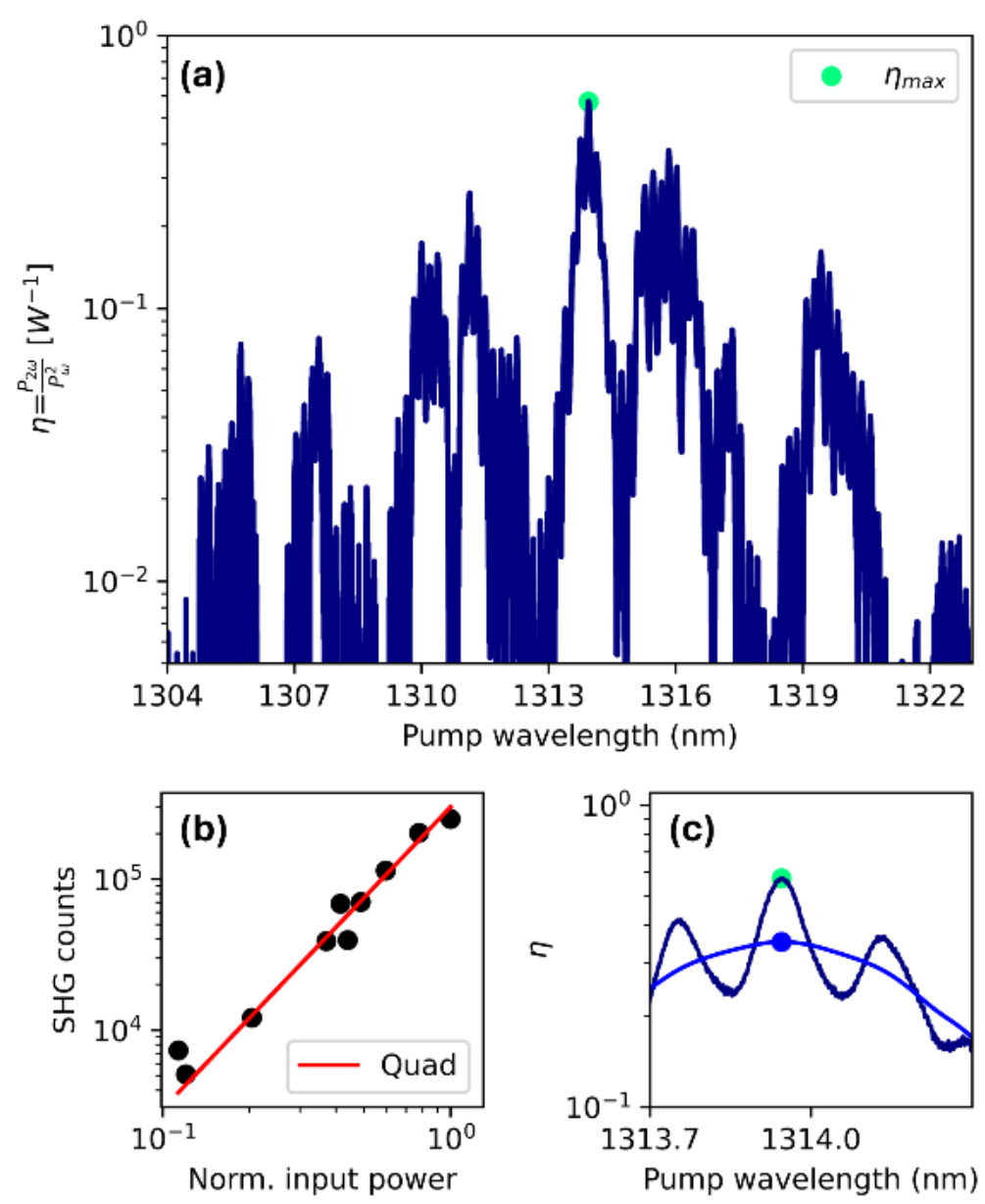

Figure 4 (a) Experimental conversion efficiency (b) Quadratic dependence of the SH power as a function of the pump power. (c) Zoomed-in view around the maximum of the experimental conversion efficiency plot.

Figure 4(a) shows $\eta$ measured over a 20 nm wavelength scan. The maximum value is recorded at 1314 nm, close to the phase matching condition designed at 1310 nm, as presented in Figure 1. The slight discrepancy is attributed to uncertainties in the waveguide dimensional measurements obtained from SEM images. Maximum conversion efficiency value of 3400% /W/cm² is measured, in the same order of magnitude than more mature platforms [23]. This value is particularly promising provided the room of improvement available on the TOP-GaP platform, especially with regard to the management of propagation losses. Fig. 4(b) shows the quadratic dependence of the measured signal as a function of the input power, confirming the experimental observation of second harmonic generation. The measurement was conducted using an alternative detection system, the StellarNet Blue-Wave spectrometer, which permits adjustable integration times, enhancing sensitivity for low-signal detection. As shown in Fig. 3(c), conversion efficiency is modulated by the contribution of Fabry-Perot interference. The free spectral range of 200 pm which is measured, closely match the calculated values of and 190 pm of $TE_{00}$ mode, respectively, with corresponding group indices of 3.46. Filtering out this interference, an average conversion efficiency is also plotted in Fig. 4(c), featuring a maximal value of 2150 %/W/cm². Injecting experimental values of the propagation losses in equation (3), the theoretical maximum conversion efficiency was calculated to be 4100 %/W/cm² in qualitative agreement with the measured one. The difference between experimental and theoretical values of the conversion efficiency can be correlated with the secondary SHG peaks observed in the vicinity of the main one and not predicted by theory as shown in Fig. S1(b). These secondary peaks are most likely due to shifts of the phase matching condition in regions of the waveguide where the resist mask has partially peeled of the GaP waveguide. Air gaps between the two GaP membranes can also locally shift the phase matching condition, producing such secondary SHG peaks.[29] The performances of TOP-GaP nanowaveguides are thus at the same order of magnitude as these of state-of-the-art PPLN integrated technologies [10]. These results represent a considerable progress compared to modal phase matching schemes previously implemented in III-V nonlinear devices, including sub-wavelength waveguides technologies [13], mainly due to the large nonlinear overlap enabled by the TOP strategy. Compared to reported OP-GaP technologies [12] however, the net gain in conversion efficiency comes only from the lower effective area of the TOP-GaP waveguides.

## 5. Conclusion

In conclusion, we have successfully demonstrated that TOP-GaP nanowaveguides can achieve modal phase matching between the $TE_{00}$ and $TM_{10}$ modes with very high second-harmonic conversion efficiency, competitive with the most mature nonlinear photonic platforms. This technology, straightforwardly transferable to any III-V semiconductor platform paves the way to ultra-high frequency conversion from the blue range with nitride to the MIR with arsenides. Future efforts will focus on reducing propagation losses in our devices through the management of the GaP/GaAs interface by polishing methods, a prerequisite to the fabrication of advanced photonic circuitry towards quantum photonic applications on chips.

### Acknowledgment

The authors acknowledge RENATECH with nanoRennes for technological support. This research was supported by "France 2030" with the French National Research agency OFCOC project (ANR-22-PEEL-0005) and the European Union (ERC-COG-2022, PANDORA, 101088331).

### Data availability

Data underlying the results presented in this paper are not publicly available at this time but may be obtained from the authors upon reasonable request.

### Disclosures

The authors declare no conflicts of interest, financial or non-financial, dealing with the present work.

### Supplementary material

See Supplement 1 for supporting content.